\documentclass{sbrt2018eng}
\usepackage{mathrsfs,amsmath}
\usepackage{amsfonts}
\usepackage[utf8]{inputenc}
\usepackage[english, brazil]{babel}

\usepackage{indentfirst,amssymb,newlfont,bm}
\usepackage[usenames,dvipsnames]{color}
\begin{document}
\title{A Note on the Shannon Entropy of Short Sequences}

\author{H. M. de Oliveira and Raydonal Ospina\\
{Statistics Department, Federal University of Pernambuco (UFPE), Recife-PE, Brazil, \\
R. Ospina is also with CAST Laboratory, E-mail: \{hmo,raydonal\}@de.ufpe.br.} }

\maketitle

\begin{abstract}
For source sequences of length $L$ symbols we proposed to use a more realistic value to the usual benchmark of number of code letters by source letters. Our idea is based on a quantifier of information fluctuation of a source, $F(U)$, which corresponds to the second central moment of the random variable that measures the information content of a source symbol. An alternative interpretation of \textit{typical sequences} is additionally provided through this approach.

\end{abstract}


\section{Introduction}
Entropy is one of the most fundamental concepts of Science \cite{0,5}. In the scope of Information Theory (IT), it is characterized in the process of coding discrete memoryless sources (DMS), through the first Shannon theorem \cite{1}, who proved it by using the law of large numbers. Source coding intends to represent symbols of an information source with as few code digits per source symbol as possible. It even seems a little odd that the entropy being associated with a statistic mean (first moment), but no parameters have been associated with the variance (second central moment) of the information produced by the symbols of the source. Interpreting the entropy $H (U) $ as the first moment of a random variable that measures the information about a symbol emitted by the source, we now take into account variations of the information content, complementary to (and associated with) entropy: the \emph{information fluctuation}, $F (U) $, which corresponds to the square root of the second central moment of the same variable. Instead of using the entropy value as a lower bound on the average letter length of the source for coded sequences, it is proposed here to adopt a limit established according to a level of significance $\alpha$ arbitrated by the coder designer. In the framework of measuring sequence entropy, an excellent reference is the work of Sch\"{u}rmann and Grassberger \cite{6}. Even so, it is assigned a point estimate, rather than an interval estimate.
\section{Fluctuation of Information}
Consider a DMS source denoted by $U$ as being a random variable which takes symbols from a finite alphabet i.e $\mathbb{A}:=\{a_k\}_{k=1}^K$ with probability of occurrence at $l$-time instant $\{p_l(a_k)\}_{k=1}^K,$ where $p_l(a_k)=P_l(U=a_k)$ for $~l=1,2,\cdots,L$ For this model of discrete source, the stationary distribution is independent of the time,  i.e., $p_l(a_k)=p(a_k)=P(U=a_k),$ and  for each symbol $a_k$ produced, the amount of information associated is
\begin{equation}
\label{eq:1}
{\cal I}(a_k)=-\log_2(p(a_k)),~a_k \in \mathbb{A} .
\end{equation}
For the sake of simplicity, let us denote $p(a_k)=p_k$. A smooth functional of the probability distribution of the source (first moment) 
is recognized as the \emph{Shannon entropy} of the DMS \cite{2}:
\begin{align}\label{entropy}
H(U):=-\sum_{k=1}^K p_k\cdot\log_2 p_k.
\end{align}
Here, we propose to use the \emph{information fluctuation} as being a compound of a functional based on the second moment of the probability distribution of the source given by
\begin{align}\label{fluctuation}
F^2(U):=\sum_{k=1}^K p_k \cdot \log_2^2 p_k-H^2(U).
\end{align}
Therefore, the fluctuation can be computed by ($F^2 \geq 0$ by Jensen's inequality):
\begin{align}\label{fluctuation formula}
F(U)=\sqrt{\sum_{k=1}^K (p_k-p_k^2) \log_2^2 p_k-\sum_{i \neq j}p_i p_j \log_2 p_i \log_2 p_j}.
\end{align}
\textbf{Definition 2.1} (degenerated sources). \textit{Let $U^*$ and $U^{**}$ be two kinds of degenerated DMS, with distributions:}\\

\indent
\textit{{\bf type I source $\bm{U^*}$:} $\{p_k\}_1^K$ where $\exists ~ k^* |p_{k^*}=1,$ and $p_k=0~(\forall ~k \neq k^*)$.}\\
\indent
\textit{{\bf type II source $\bm{U^{**}}$:} $\{p_k\}_1^K,$ where $p_k=\frac{1}{J},$
	 $\forall ~k \in I \subset \{1,2,\cdots,K\}$ and $J=||I||$. Here, the $||\cdot||$ indicates the cardinality of set.} 
$\square$\\

\begin{figure}[!ht]
\centering{\includegraphics[width=75mm]{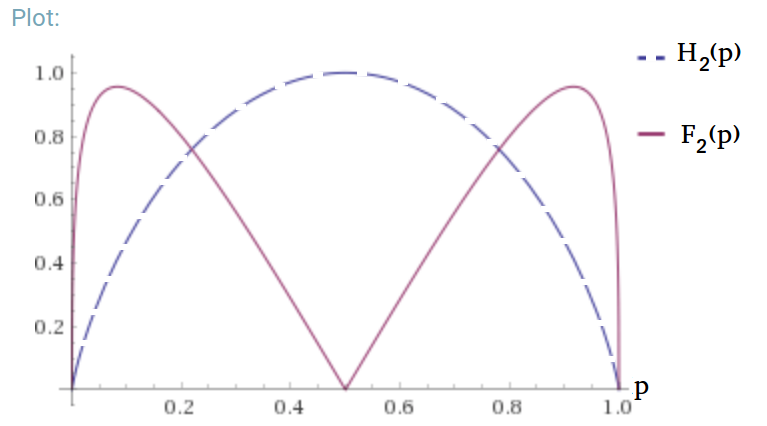}}
\caption{\label{mainfig}Binary fluctuation information $F_2(p)$ expressed in Shannons.}
\end{figure}

\begin{figure}[!ht]
\centering{\includegraphics[width=75mm,height=50mm]{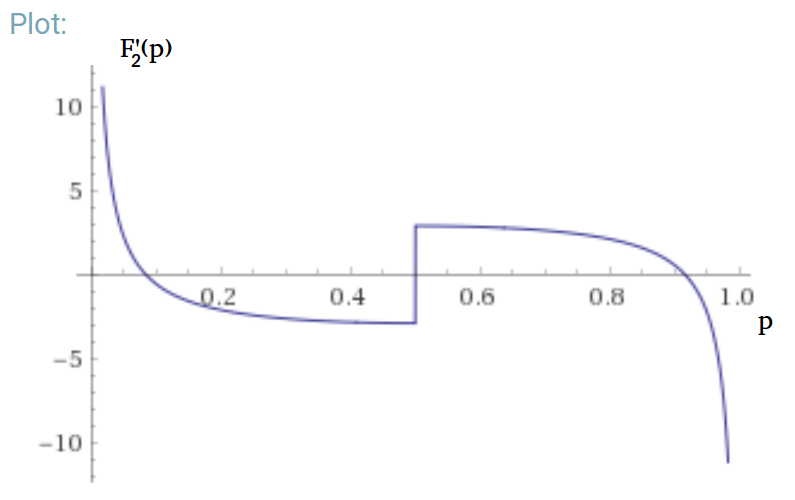}}
\caption{\label{deriva}Derivative of the binary fluctuation $F_2(p)$.}
\end{figure}

\noindent
\textbf{Proposition 2.1}. \textit{The fluctuation of information is null if and only if the source is degenerated, i.e.,} $F(U)=0 \Leftrightarrow U^*$ \textit{or} $U^{**}$.\\
\textbf{proof}. ($\Leftarrow$) $F(U^*)=0$ follows directly from $p_k=p_k^2~ (\forall k)$ and $p_i p_j=0~ \forall i \neq j$, in this case.
For type II sources, there are $J >1$ identical terms in the first summation and $J^2-J$ identical terms in the second summation in Equation  (\ref{fluctuation formula}). Therefore
\begin{subequations}
\begin{align}
\sum_{k=1}^J (p_k-p_k^2) \log_2^2 p_k &= J \Big[\frac{1}{J} - \frac{1}{J^2}\Big] \log^2 \frac{1}{J},\\
\sum_{i \neq j} p_ip_j \log_2 p_i \log p_j &= (J^2-J) \cdot \frac{1}{J^2}\Big(\log \frac{1}{J}\Big)^2.
\end{align}
\end{subequations}
($\Rightarrow$) Without loss of generality, consider that $\{p_k\}_1^K$ has been sorted in descending order, $p_1 \leq p_2 \leq \cdots \leq p_K$.
This means that if $p_k>0$, since the terms $p_k=0$ do not count. Thus, the possible solutions are of the type
$p_k =\frac{1}{J} \text{ for } k=1,2,\cdots,J \leq K$ and $p_k = 0,  ~\forall k>J$.\\ 
Two trivial solutions are:\\
\indent 
\begin{itemize}
\item $p_1 = 1 \text{ and } p_k=0, \text{ for } k>1 ~(J=1),$
\item $p_k = 1/J, \text{ for } k=1,2,\cdots ~(J=K)$, 
\end{itemize} concluding the proof. \\
\\
Although $U^*$-sources can be included as a particular case of $U^{**}$-sources (by assuming $J =||I||=1$), we have decided to define them as different types of sources. 
As an illustration, consider now a binary DMS source where the output is either a 0 with probability $p$ or 1 with a probability $1-p.$
For this case the, the quantifiers given in (\ref{entropy}) and  (\ref{fluctuation}) simplifies for the following two expressions, respectively:
\begin{align}\label{binary entropy}
H_2(p):=-p \log_2 p - q \log_2 q,
\end{align} 
\begin{align}\label{binary fluctuation}
F_2(p):=\sqrt{p q \left ( \log_2 p -\log_2q  \right )^2},
\end{align}
where $q:=1-p$.
The shape of the binary entropy $H_2(\cdot)$ is well-known \cite{2}, and we sketch $F_2(\cdot)$ in the Fig.~\ref{mainfig}.
There are five attributions to $p$ worth to mention: $p \in \{0, 0.0832217\cdots, 0.5, 0.9167783\cdots,1\}$. Three limit cases have no variance (binary degenerated sources): $p=0$ and $p=1$ (source transmits just the same symbol and has null entropy), and $p=0.5$. This last case, somewhat surprisingly, with equiprobable sequences and maximum entropy, has no variance in the amount of information: whatever the length of the sequence of symbols of the source, the average information per source letter is \emph{exactly} equal to the entropy. However, the points of maximum variability on the information content seem to be less known in the literature. We start by investigating the behavior of $\frac{\mathrm{d} F_2(p)}{\mathrm{d} p}$. We find that
\begin{align}\label{derivative of fluctuation}
\frac{\mathrm{d} F_2(p)}{\mathrm{d} p}=\frac{\log (\frac{1}{p}-1).\left \{ (1-2p) \tanh^{-1}(1-2p)-1 \right \} }{\sqrt{-p(1-p) \log^2 (\frac{1}{p}-1)}}.
\end{align}
A plot of the derivative of $F_2(p)$ is shown in Figure \ref{deriva}. Three points have infinite derivative values: exactly the limiting cases aforementioned. The two critical points occurs when $\tanh \left ( \frac{1}{1-2p} \right )=1-2p$, and the numerical solution of $\tanh \frac{1}{x}=x$ furnishes $x^*=\pm 0.833556559600964698 \cdots$, yielding:
$p^* \approx \frac{1}{2} [1 \pm 0.833557]$. The value of the derivative is about 0.956137 at the two maximum points. The \emph{saltus} at $p=0.5$ is circa $5.77078$.\\
We also calculate the intersection points of the two curves in Fig. \ref{mainfig}, where $H_2(p)=F_2(p)$. In the region between the two points, the standard deviation of the entropy is smaller than the entropy itself. We agreed to use this range of $p$ values to characterize binary DMS with low entropy variability.\\

\begin{figure}[!ht]
\centering{\includegraphics[width=68mm,height=45mm]{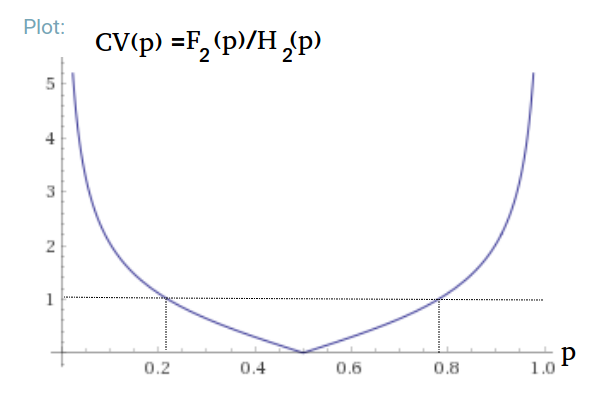}}
\caption{\label{CV}Coefficient of variation of entropy: $CV:=100 \frac{F_2(p)}{H_2(p)}$.}
\end{figure}

Using the coefficient of variation \cite{3}, $CV:=100 \frac{F_2(p)}{H_2(p)}$, we plot Fig. \ref{CV}.
\\
From analysis the behavior of Fig. \ref{CV} we obtain the next definition. \\
\textbf{Definition 2.2}. \textit{A binary memoryless source is called a low entropy variability DMS if and only if $0.21907592 \leq p \leq 0.78092407$.} $\square$\\
\section{Entropy of Finite Sequences: Statistical Evaluation}

Let $\{x_l\}$ be a sequence of observed counts of symbols in the alphabet in a DMS sample of size $L$ and $\widehat p_k=m_k/L$ as being the sample relative frequency of the $k$th symbol $a_k$ (here, $m_k$ is the number of symbols $a_k$ in the sample).  The most commonly nonparametric estimation  of  $H(U)$ is given by the plug-in estimator 
\begin{equation*}
\label{eq:3}
\widehat H(U):=-\sum_{k=1}^K \widehat p_k\cdot\log_2 \widehat p_k
\end{equation*}
and  has been a subject of much research for many decades (See \cite{7}--\cite{11}).
In the same way, we propose to use as plugin estimator of information fluctuation the statistic
\begin{equation}
\label{eq:4}
\widehat F^2(U):=\sum_{k=1}^K \widehat p_k \cdot \log_2^2 \widehat p_k-\widehat H^2(U).
\end{equation}
Using Taylor expansion of the functions $\widehat H(U)$ and $\widehat F^2(U)$ about the point $(p_1, \ldots, p_L)$ and confining ourselves to derivatives of fourth order is possible to proof that the estimators 
$\widehat H(U),$ and $\widehat F^2(U)$ are asymptotically unbiased and consistent estimators of entropy $H(U)$ and information fluctuation $F^2(u),$  respectively.
Basharin \cite{8} shown that  if $L$ is large and given the independence of the informations, the central limit theorem can be used to establish that  $\widehat{H}(U)$ converges to the normal distribution
\begin{align}
\widehat{H}(U) \sim {\cal N}(H(U), \frac{F^2(U)}{L}).
\end{align}
Using a similar heuristic used  in \cite{8}, is possible to note that $\widehat F^2(U)$ is in the fact an asymptotic estimator of he information fluctuation, i.e. $\mathbb{V}ar(\widehat{H}(U))$ is the variance of entropy. Consequently the quantity 
\begin{align}
\frac{(L-1)\widehat{F}^2(U)}{F^2(U)} \sim \chi^2 (L-1),
\end{align}
where $\chi^2$ indicates converges  to the Chi-square distribution.

Set a significance level $\alpha$, assuming the population variance, $F^2(U)$, is known, one can calculate a (right) confidence interval setting  \cite{3}
\begin{align}
\widehat{H}(U)+z_{\alpha} \frac{F(U)}{\sqrt{L}},
\end{align}
where $z_{\alpha}=\phi^{-1}(\alpha)$ is the $\alpha$-quantile of the Normal distribution, where $\phi(z):=\frac{1}{\sqrt{2 \pi}} \int_{-\infty}^z e^{-\zeta^2/2} d \zeta.$
Naturally, for short sequences is more appropriate to use the robust $t$-Student statistics:
\begin{align}\label{practical entropy}
H_{practical}:= \widehat{H}(U)+t_{(\alpha;L-1)} \frac{\widehat{F}(U)}{\sqrt{L}},
\end{align}
where $t_{(\alpha, L-1)}$ is the critical value with $L-1$ degrees of freedom  of $t$-Student distribution \cite{3}.
It is worth remembering that the values of $H(U)$ and $F(U)$ are not dissociated. In the binary case, for example, each source with $Ber(p)$ distributed symbols (known), there are pairs $(H, F)$ given by Equations (\ref{binary entropy}) and (\ref{binary fluctuation}), respectively. Set a confidence level, $1-\alpha$, say 95\%. For a large number of sequence outcomes for the source (all with length $L$), the sample entropy value does not exceed the value given in Equation (\ref{practical entropy}) in 95\% of cases. We can thus expect to code the output with $H_{practical}$ bits per source symbol, instead of using the fundamental limit $H(U)$, which can be asymptotically achieved when the length of the source sequences grows indefinitely. Instead of adopting the conventional figure $\eta:=H/\bar{L}$, where $\bar{L}$ is the average number of bits per source letter, we propose to assess the efficiency of a source coding run for a particular given sequence (especially in the cases of short sequences and high entropy variability DMS), at a significance level $\alpha$, according to
\begin{equation}
\label{efficiency}
\displaystyle{
\eta_{\alpha} := \frac{\widehat{H}(U)+t_{(\alpha, L-1)}\widehat{F}(U)/\sqrt{L}}{\bar{L}}.}
\end{equation}

\section{Atypical Sequences}
Let us now shift the focus to long sequences. A statistical interpretation for ``atypical sequences'' can be formulated in this scenario. All sequences produced from the source that result in a sample entropy within the confidence interval 
\begin{align}\label{IC}
\Big[\widehat{H}(U)-z_{\alpha/2} \frac{\widehat{F}(U)}{\sqrt{L}}, \widehat{H}(U)+z_{\alpha/2} \frac{\widehat{F}(U)}{\sqrt{L}}\Big]
\end{align}
may are considered to be typical. On the other side, the atypical sequences are those corresponding to the level of significance adopted. Now, set an arbitrary value $\epsilon>0$. Any sequence (a source sample) with length $L$ symbols for which the sample entropy is outside the range $[H-\epsilon, H +\epsilon]$ is called an $\epsilon$-atypical sequence. What is the level $\alpha$ of significance at which the confidence interval (Eq. (\ref{IC})) for the entropy coincides with this interval? The relationship
\begin{align}\label{atypical}
\phi^{-1}\Big(1-\frac{\alpha}{2}\Big)=\frac{\epsilon \sqrt{L}}{F(U)}
\end{align}
should be imposed. Here, the significance level is interpreted as the probability of occurrence of an atypical sequence, i.e., $P(\text{atypical sequence}) = \alpha$. Therefore,
\begin{align}
\lim_{L \to \infty} P(\text{atypical sequence})=\lim_{L \to \infty} 2\left [ 1-\phi \left ( \frac{\epsilon \sqrt{L}}{F(U)} \right ) \right ]=0.
\end{align}
Thus, the significance level becomes arbitrarily small as the $L$ grows. Typical sequences (in the sense that the sample entropy is $\epsilon$-close the entropy of the source) constitute virtually 100\% of the sequences, provided that the sequence length $L$ is sufficiently large. This statistical interpretation is immediately understandable. \\
Another key concept of IT is the asymptotic equipartition property (AEP). The extension $U^L$ of the source $U$ consider $L$-grams as the new symbols, it has an alphabet $\{\underline u_i \}_{i=1}^{K^L}$ with probabilities $\{P( \underline u_i) \}_{1}^{K^L}$. More than 
$P \left [ \lim_{L \to \infty} \widehat{H}(U)=H(U) \right ]=1$, it is shown that $\epsilon$-typical sequences hold \cite{4}
\begin{align}\label{AEP}
2^{-L(H(U)+\epsilon)} \leq P( \underline u_i) \leq 2^{-L(H(U)-\epsilon)}.
\end{align}
Therefore, $P( \underline u_i) \approx 2^{-L\cdot H(U)}$ (constant) for $i \in  \mathbb{T} \subset \{1,2,\cdots, K^L\}$, the set of typical sequences, with $\parallel \mathbb{T} \parallel = 2^{L\cdot H(U)}$. By definition of degenerated sources, we see that large extensions of a DMS becomes type II asymptotically degenerated. Its entropy becomes exactly $L\cdot H(U)$, with essentially no information fluctuation.

\section{Concluding Remarks}

In addition to proposing the calculation of a more ``realistic'' estimate for entropy, now depending on the length of the sequence by the source, a naive and didactic interpretation on ``typical sequences'' is presented. The idea of introducing a new parameter for a DMS, namely the information fluctuation, seems to be as basic as the concept of entropy itself. For short sequences is not the entropy of the source that should be used to calculate the efficiency of source coders, but rather in terms of the sample entropy. Another potential application is in the assessment of randomness of sequences. Some open questions remains: There exists always a maximum fluctuation in the sample entropy for some particular (non-binary) probability assignment?

\section*{Acknowledgements}
This work has been supported by Statistics Department UFPE, Brazil.

\end{document}